\documentclass[aps,pra,twocolumn,showpacs,10pt]{revtex4-1}
\usepackage{amsmath,amssymb,graphicx}
\begin{document}

\title{Phase- and frequency-controlled interference nonlinear optics in superconducting circuits
}

\author{Hai-Chao Li}\email{lhc2007@hust.edu.cn}
\affiliation{School of Physics,
Huazhong University of Science and Technology, \\
Wuhan, Hubei 430074, People's Republic of China}

\author{Hai-Yang Zhang}
\author{Qing He}
\author{Guo-Qin Ge}\email{gqge@hust.edu.cn}
\affiliation{School of Physics,
Huazhong University of Science and Technology, \\
Wuhan, Hubei 430074, People's Republic of China}

\begin{abstract}
We present a new type of phase- and frequency-sensitive amplification and attenuation in a cyclically driven three-level superconducting Josephson system. Different from the previous linear theory of pure phase-sensitive amplification, a new physical mechanism$-$combined action of nonlinear wave mixing and wave interference$-$is developed and leads to not only amplification but also attenuation. This is referred to as interference nonlinear optics. Our results show the sudden output signal transition from large gain to deep suppression by tuning the relative phase and in this case the system can act as a phase-controlled amplitude modulator. We also show the continuous change from output enhancement to attenuation by adjusting the driving-field frequency and in this situation the system behaves as a frequency-controlled amplitude modulator. Our study opens up a new perspective for its widespread applications in quantum information science.
\end{abstract}


\maketitle
Solid-state superconducting circuits~\cite{Blais,Wallraff} based on Josephson junctions are versatile quantum mechanical systems in which superconducting quantum qubits~\cite{You1,Clarke,Xiang}$-$artificial multi-level atoms$-$can be tuned and controlled unprecedentedly by external gate voltage and magnetic flux. In a series of theoretical and experimental works~\cite{Rebi,HLi,Grajcar,Sun,Hoi}, such circuit architectures successfully have been used to produce numerous quantum optical phenomena in the microwave frequency domain, which opens up the interesting realm for studying circuit quantum electrodynamics (QED)~\cite{You}. Especially, some novel or previously unproved physical phenomena have been demonstrated and observed in circuit QED platform, such as ultrastrong coupling regime~\cite{Niemczyk}, collapse and revival of a coherent state with single-photon Kerr regime~\cite{Kirchmair}, the dynamical Casimir effect~\cite{Wilson}. All these progresses have intensely stimulated the research for further exploring fundamental quantum physics as well as potential applications in quantum information processing in the circuit QED architecture providing an artificial medium with engineered atom-field interaction.

It is well known that exploiting the second-order nonlinearity in cavity QED is severely restrained due to the presence of selection rules based on the inversion symmetry of potential energy in atomic systems. In contrast, for artificially designed superconducting quantum qubit selection rules do not work when the qubit's inversion symmetry is broken~\cite{Liu,Manucharyan}. For instance, superconducting fluxonium qubit can have a cyclic $\triangle$-type three-level structure, which is beyond selection rules under the electric-dipole approximation and has been demonstrated in experiment. The absence of selection rules is an important mechanism for many interesting quantum physical phenomena~\cite{Youjq,Youl,Liuy}. In our previous work coexistence of three-wave, four-wave, and five-wave mixing processes has been shown using the $\triangle$-type superconducting system~\cite{HcLi}. Also, the absence of selection rules explains the surprisingly large dispersive shifts observed in fluxonium experiments and leads to the prediction of a two-photon vacuum Rabi splitting~\cite{Zhu}. Moreover, the preparation of nonclassical microwave states~\cite{Zhao} via longitudinal-coupling-induced multi-photon processes~\cite{Liu2} has been demonstrated in a driven inversion-symmetry-broken superconducting quantum system.

In this paper, we present a new type of phase- and frequency-sensitive amplification and attenuation in the microwave frequency domain in terms of superconducting quantum circuits.
Here using a cyclic $\triangle$-type three-level artificial system driven by three incoming waves, two reverse three-wave mixing processes, sum- and difference-frequency generation with second-order nonlinearity, can exist simultaneously. And by arranging two available matched conditions among three incident tones, two sets of wave interferences between the incoming signals and the generated signals appear and play a crucial role in amplifying or attenuating respective output fields. Owing to the output signals being sensitive to the relative phase of the incident fields or the driving-field frequency, the superconducting system can act as a phase- or frequency-controlled amplitude modulator, an important device having a number of potential applications in quantum information processing. The mentioned phase- and frequency-sensitive amplification and attenuation originate from a new physical mechanism, i.e., combined action of nonlinear wave mixing and wave interference. This obviously discriminates our present scheme from the linear theory of pure phase-sensitive amplification~\cite{Scully,Ansari} in the three-level cascade-type atomic system where the transition between the top and bottom levels is dipole forbidden. Moreover, although multi-wave mixing has been studied in atomic systems~\cite{Nie,Zhang1,Lic}, intermixing only between nonlinear optical processes is involved and leads to enhancement and suppression of the generated waves instead of the incident waves.

\begin{figure}[htbp]
\centerline{
\includegraphics[width=0.6\columnwidth]{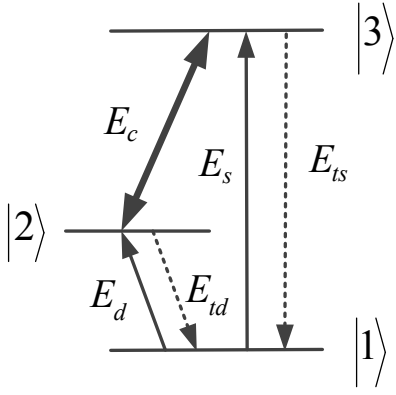}}
\caption{Schematic of a $\triangle$-type three-level superconducting artificial system interacting with three incoming waves (solid lines). A weak driving field $E_{d}$ couples the $|1\rangle$ $\leftrightarrow$ $|2\rangle$ transition and another weak signal field $E_{s}$ drives the $|1\rangle$ $\leftrightarrow$ $|3\rangle$ transition. Meanwhile, levels $|2\rangle$ and $|3\rangle$ are connected by a strong coupling field $E_{c}$. In this atom-field interaction configuration, two reverse three-wave mixing processes, sum-frequency $E_{ts}$ and difference-frequency $E_{td}$ generation (dashed lines), can exist simultaneously.
}
\label{fig.1}
\end{figure}

Let us consider a cyclic three-level superconducting artificial system interacting with three microwave fields, as depicted in Fig.~\ref{fig.1}. Two weak incoming waves, a driving field $E_{d}$ ($\omega_{d}, {\rm{\textbf{k}}}_{d}$ and  Rabi frequency $\Omega_{d}$) and a signal field $E_{s}$ ($\omega_{s}, {\rm{\textbf{k}}}_{s}$ and Rabi frequency $\Omega_{s}$), couple the transitions $|1\rangle\leftrightarrow|2\rangle$ and $|1\rangle\leftrightarrow|3\rangle$, respectively. Meanwhile, a strong control field $E_{c}$ ($\omega_{c}, {\rm{\textbf{k}}}_{c}$, and Rabi frequency $\Omega_{c}$) connects the upper transition $|2\rangle$ to $|3\rangle$. According to the above atom-field interaction configuration shown in Fig.~\ref{fig.1}, there are two coexisting reverse three-wave mixing processes with second-order nonlinearity via transitions in different
branches. To be specific, a three-wave mixing sum-frequency signal $E_{ts}$ with phase-matching condition ${\rm{\textbf{k}}}_{ts}={\rm{\textbf{k}}}_{d}+{\rm{\textbf{k}}}_{c}$ is generated via the Liouville pathway $\rho^{(0)}_{11}\rightarrow\rho^{(1)}_{21}\rightarrow\rho^{(2)}_{31}$ and another three-wave mixing difference-frequency signal $E_{td}$  with phase-matching condition ${\rm{\textbf{k}}}_{td}={\rm{\textbf{k}}}_{s}-{\rm{\textbf{k}}}_{c}$ is generated via the Liouville pathway $\rho^{(0)}_{11}\rightarrow\rho^{(1)}_{31}\rightarrow\rho^{(2)}_{21}$. It should be pointed out that in our three-wave mixing scheme the following frequency relation among three incident microwaves is satisfied, i.e., $\omega_{s}=\omega_{d}+\omega_{c}$. Thus the sum-frequency $E_{ts}$ (difference-frequency $E_{td}$) generation has the same traveling frequency with the signal $E_{s}$ (driving $E_{d}$) field.

In the interaction picture, the Hamiltonian of the artificial $\triangle$-type system interacting with three incoming waves and two generated waves under the rotating-wave approximation is given by ($\hbar=1$)
\begin{align}
H&=\Delta_{d}\sigma_{22}+\Delta_{d}\sigma_{33}-\frac{1}{2}(\Omega_{d}\sigma_{21}\nonumber \\
&\quad+\Omega_{td}\sigma_{21}+\Omega_{c}\sigma_{32}+\Omega_{s}\sigma_{31}+\Omega_{ts}\sigma_{31}+{\rm{H.c.}}),
\end{align}
where $\sigma_{ij}=|i\rangle\langle j|$  denotes the atomic transition operator, $\Omega_{td}$ ($\Omega_{ts}$) is the Rabi frequency of difference-frequency (sum-frequency) field, $\Delta_{d}=\omega_{21}-\omega_{d}$ is the detuning of the driving field. The control field frequency $\omega_{c}$ is assumed to be resonant with the energy spacing $\omega_{32}$. Including the relaxation and dephasing processes, the
evolution of dynamics for the superconducting system can be described by a Lindblad-type master equation
\begin{align}
\frac{d\rho}{dt}&=-i[H, \rho]+\frac{1}{2}\sum_{j=2}^{3}\gamma_{\phi j}(2\sigma_{jj}\rho\sigma_{jj}-\sigma_{jj}\rho-\rho\sigma_{jj})\nonumber \\
&\quad+\frac{1}{2}\sum_{i<j}\gamma_{ij}(2\sigma_{ij}\rho\sigma_{ji}-\sigma_{jj}\rho-\rho\sigma_{jj}).
\end{align}
Here $\gamma_{ij}$ represents the relaxation rate between the levels $|i\rangle$ and $|j\rangle$, and $\gamma_{\phi j}$ denotes the pure dephasing rate for level $|j\rangle$ and it is negligible for a superconducting
fluxonium system in a wide range of flux around a degeneracy point~\cite{Manucharyan,Manucharyan1}.

We now demonstrate the scheme for controllable phase- and frequency-sensitive microwave amplification and attenuation via the combined action of nonlinear three-wave mixing and wave interference in the presence of a strong control field limit in which linear absorption can be greatly suppressed while nonlinear optical processes are resonantly enhanced~\cite{Li1,Kang}.
It is well known that linear and nonlinear polarizations for a quantum multi-level system depend upon first-order and high-order off-diagonal density matrix elements which can be obtained by a formal perturbation expansion. Assuming that the initial population of the considered system is prepared in the ground state $|1\rangle$,  the steady-state solutions of the matrix elements associated with the two transition paths $|1\rangle\leftrightarrow|2\rangle$ and $|1\rangle\leftrightarrow|3\rangle$ are expressed as
\begin{subequations}
\begin{align}
\rho^{(1)}_{21}&=\frac{i(\Omega_{d}+\Omega_{td})(\Gamma_{31}+i\Delta_{d})}{2\xi},\\
\rho^{(1)}_{31}&=\frac{i(\Omega_{s}+\Omega_{ts})(\Gamma_{21}+i\Delta_{d})}{2\xi},\\
\rho^{(2)}_{21}&=\frac{i^{2}\Omega_{c}^\ast\Omega_{s}}{4\xi}+\frac{i^{2}\Omega_{c}^\ast\Omega_{ts}}{4\xi},\\
\rho^{(2)}_{31}&=\frac{i^{2}\Omega_{c}\Omega_{d}}{4\xi}+\frac{i^{2}\Omega_{c}\Omega_{td}}{4\xi},
\end{align}
\end{subequations}
where $\xi=(\Gamma_{21}+i\Delta_{d})(\Gamma_{31}+i\Delta_{d})+|\Omega_{c}|^{2}/4$, $\Gamma_{31}=\frac{1}{2}(\gamma_{13}+\gamma_{23}+\gamma_{\phi3})$ and $\Gamma_{21}=\frac{1}{2}(\gamma_{12}+\gamma_{\phi2})$. Equations (3a) and (3b) describe the linear susceptibilities, which control the absorption and dispersion characteristics of the incident driving, signal, the generated sum- and difference-frequency fields. The first terms in Eqs. (3c) and (3d) illustrate the difference- and sum-frequency generation with second-order nonlinearity, and the second terms indicate the backward nonlinear processes of two generated three-wave mixing fields.

We emphasize that an interesting and important phenomenon$-$interference between the incoming waves and the generated waves$-$occurs in our project when another condition ${\rm{\textbf{k}}}_{s}={\rm{\textbf{k}}}_{d}+{\rm{\textbf{k}}}_{c}$ in among three incoming waves is satisfied synchronously. In that case the sum-frequency and signal (difference-frequency and driving) fields propagate along the same direction ${\rm{\textbf{k}}}_{s}$ (${\rm{\textbf{k}}}_{d}$) , and as a result, they are indistinguishable and the total output $E_{s}^{tot}$ ($E_{d}^{tot}$) can be considered as a coherent superposition of these two signals, i.e., $E_{s}^{tot}=E_{s}+E_{ts}$ ($E_{d}^{tot}=E_{d}+E_{td}$).
Using the slowly varying amplitude approximation~\cite{Shen} and solving two sets of coupled wave equations for the fields $E_{d}$ and $E_{ts}$ and  the fields $E_{s}$ and $E_{td}$, the total output signals $E_{s}^{tot}$ and $E_{d}^{tot}$ are obtained
\begin{widetext}
\begin{align}
E_{s}^{tot}/E_{s0}&=(E_{ts}+E_{s})/E_{s0} \nonumber \\&
= G\cos\left(\frac{FZ}{4\xi}\right)
-\frac{i\Delta_{d}G(\Gamma_{31}/\Gamma_{21}-1)}{F}\sin\left(\frac{FZ}{4\xi}\right)
-\frac{ie^{-i(\phi_{d}+\phi_{c}-\phi_{s})}G|\Omega_{d0}||\Omega_{c}|/|\Omega_{s0}|}{\sqrt{\Delta_{d}^{2}(1-\Gamma_{21}/\Gamma_{31})^2+|\Omega_{c}|^2\Gamma_{21}/\Gamma_{31}}}\sin\left(\frac{FZ}{4\xi}\right),\\
E_{d}^{tot}/E_{d0}&=(E_{td}+E_{d})/E_{d0} \nonumber \\&
=G\cos\left(\frac{FZ}{4\xi}\right)
+\frac{i\Delta_{d}G(\Gamma_{31}/\Gamma_{21}-1)}{F}\sin\left(\frac{FZ}{4\xi}\right)
-\frac{i e^{-i(\phi_{s}-\phi_{c}-\phi_{d})}G|\Omega_{s0}||\Omega_{c}|/|\Omega_{d0}|}{F}\sin\left(\frac{FZ}{4\xi}\right),
\end{align}
\end{widetext}
where $F=\sqrt{\Delta_{d}^{2}(1-\Gamma_{31}/\Gamma_{21})^2+\Gamma_{31}/\Gamma_{21}|\Omega_{c}|^2}$, $G=\rm{exp}[-\Gamma_{31}\rm{Z}/(2\xi)-i\Delta_{d}(1+\Gamma_{31}/\Gamma_{21})\rm{Z}/(4\xi)]$, $\rm{Z}=\kappa_{12}z$ is the effective propagation distance, $\kappa_{ij}$ is a propagation constant, and we assume $\kappa_{12}\Gamma_{31}=\kappa_{13}\Gamma_{21}$ for simplicity in calculation. Here we treat the Rabi frequencies $\Omega_{d}$, $\Omega_{c}$ and $\Omega_{s}$ as complex parameters: $\Omega_{d}=|\Omega_{d}|\rm{e}^{-i\phi_{d}}$, $\Omega_{c}=|\Omega_{c}|\rm{e}^{-i\phi_{c}}$ and $\Omega_{s}=|\Omega_{s}|\rm{e}^{-i\phi_{s}}$, where $\phi_{d}$, $\phi_{c}$ and $\phi_{s}$ are the phases of the driving, control and signal fields, respectively. The first two terms in Eqs. (4) and (5) describe the evolutions of the signal and driving fields while the third terms dominate propagation dynamics of the sum- and difference-frequency fields. Clearly, the output amplitudes $E_{s}^{tot}$ and $E_{d}^{tot}$ are sensitive to the relative phase $\phi=\phi_{d}+\phi_{c}-\phi_{s}$ and interference effects between the corresponding incoming and generated waves play an essential role in amplification and attenuation of the total output signals.

\begin{figure}[t]
\centerline{
\includegraphics[width=1.1\columnwidth]{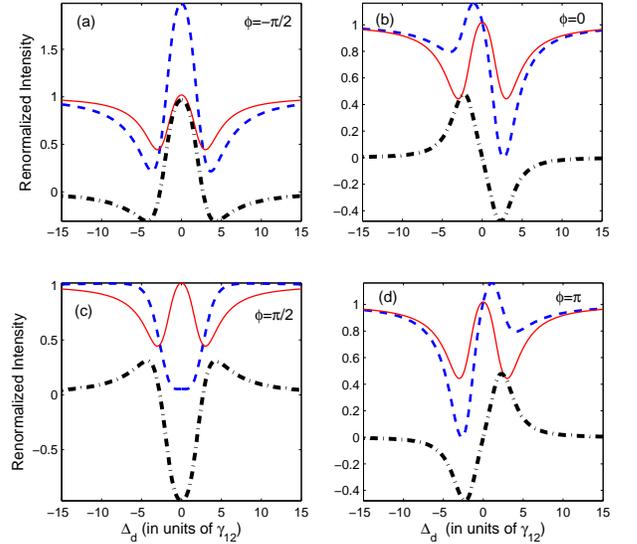}}
\caption{(Color online) Renormalized output signal intensity $|E_{s}^{tot}/E_{s0}|^{2}$ (blue dashed line), the sum $|E_{ts}/E_{s0}|^{2}+|E_{s}/E_{s0}|^{2}$ (red solid line) and the interference term $|E_{s}^{tot}/E_{s0}|^{2}-|E_{ts}/E_{s0}|^{2}-|E_{s} /E_{s0}|^{2}$ (black dash-dotted line) as a function of the driving detuning $\Delta_{d}$ for various values of the relative phase $\phi$: (a) $\phi=-\pi/2$, (b) $\phi=0$, (c) $\phi=\pi/2$ and (d) $\phi=\pi$. The other parameters are $\gamma_{13}=3\gamma_{12}$, $\gamma_{23}=\gamma_{12}$, $Z=\gamma_{12}$, $|\Omega_{d0}|/|\Omega_{s0}|=1$, and $\Omega_{c}=5\gamma_{12}$.
}
\label{fig.2}
\end{figure}

\begin{figure}[htbp]
\centerline{
\includegraphics[width=1.1\columnwidth]{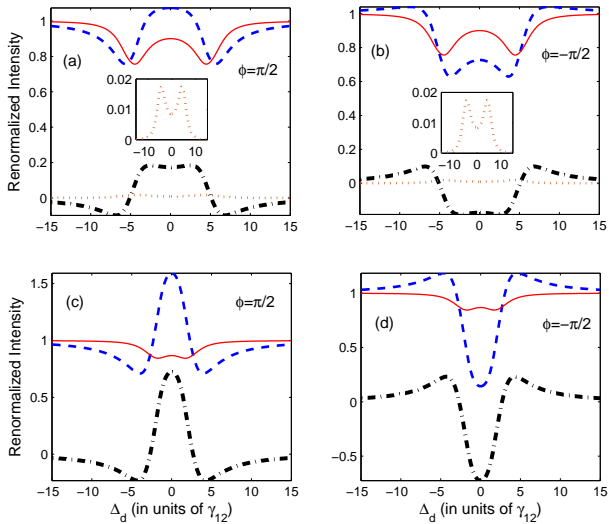}}
\caption{(Color online) Renormalized output signal intensity $|E_{d}^{tot}/E_{d0}|^{2}$ (blue dashed line), the sum $|E_{td}/E_{d0}|^{2}+|E_{d}/E_{d0}|^{2}$(red solid line) and the interference term $|E_{d}^{tot}/E_{d0}|^{2}-|E_{td}/E_{d0}|^{2}-|E_{d} /E_{d0}|^{2}$ (black dash-dotted line) versus the driving detuning $\Delta_{d}$. The inset denotes the intensity of the generated difference-frequency field $E_{td}$. (a) $\phi=\pi/2$ and $|\Omega_{s0}|/|\Omega_{d0}|=1$ , (b) $\phi=-\pi/2$ and $|\Omega_{s0}|/|\Omega_{d0}|=1$, (c) $\phi=\pi/2$ and $|\Omega_{s0}|/|\Omega_{d0}|=3$ and (d) $\phi=-\pi/2$ and $|\Omega_{s0}|/|\Omega_{d0}|=3$. The other parameters are $\gamma_{13}=3\gamma_{12}$, $\gamma_{23}=\gamma_{12}$, $Z=\gamma_{12}$, and $\Omega_{c}=10\gamma_{12}$.
}
\label{fig.3}
\end{figure}

Figure~\ref{fig.2} presents the evolutions of $|E_{s}^{tot}/E_{s0}|^{2}$, $|E_{ts}/E_{s0}|^{2}+|E_{s}/E_{s0}|^{2}$ and the interference term $|E_{s}^{tot}/E_{s0}|^{2}-|E_{ts}/E_{s0}|^{2}-|E_{s}/E_{s0}|^{2}$ as a function of the driving detuning $\Delta_{d}$ for various
values of the relative phase $\phi$ according to Eq.~(4). In Fig.~\ref{fig.2}(a) with $\phi=-\pi/2$, the total output intensity $|E_{s}^{tot}/E_{s0}|^{2}$ is larger than one and the incident signal field $E_{s}$ is amplified after passing through the artificial medium. The interference intensity is all but equal to value of the sum $|E_{ts}/E_{s0}|^{2}+|E_{s}/E_{s0}|^{2}$ in the vicinity of resonant point and the physical mechanism responsible for realizing such output gain is strong constructive interference between the incoming wave $E_{s}$ and the generated wave $E_{ts}$. Contrarily, in Fig.~\ref{fig.2}(c) with $\phi=\pi/2$ the output signal almost reduces to zero at the resonant point as the result of the strong destructive interference. Thus we achieve the sudden output signal transition from large gain to deep suppression by tuning the relative phase and in this case the system can act as a phase-controlled amplitude modulator. In addition to the pure constructive or destructive interference, the mixture of these two interferences can exist effectively.  Concretely, constructive interference in the blue-detuned region and destructive interference in the red-detuned region are shown in Fig.~\ref{fig.2}(b), and the opposite case occurs in Fig.~\ref{fig.2}(d). The corresponding results are that both output amplification and attenuation can be obtained by adjusting the detuning $\Delta_{d}$ and in this situation the system behave as a frequency-controlled amplitude modulator.

Compared with Eqs. (4) and (5), we find the output amplitude $E_{d}^{tot}$ can have the similar behaviours with $E_{s}^{tot}$ in the nonlinear system. In Fig.~\ref{fig.3}, we plot the output image of $|E_{d}^{tot}/E_{d0}|^{2}$ versus $\Delta_{d}$ for two typical relative phase values $\phi=\pi/2$ and $\phi=-\pi/2$. Because the intensity of the generated difference-frequency field $E_{td}$ (inset in Fig.~\ref{fig.3}) is far smaller than that of the sum-frequency field $E_{ts}$ for the same initial ratio $|\Omega_{s0}|/|\Omega_{d0}|=1$, to a certain extent, the interference effects are still important for signal enhancement and attenuation, but are suppressed greatly, as depicted in Figs.~\ref{fig.3}(a) and~\ref{fig.3}(b).By increasing the initial ratio $|\Omega_{s0}|/|\Omega_{d0}|$, the generated field $E_{td}$ is enhanced and interference effects are strengthened. Subsequently, we retrieve the amplification and attenuation of the output singal $E_{d}^{tot}$ in a large range shown in Figs.~\ref{fig.3}(c) and~\ref{fig.3}(d). So the system can also serve as a phase-controlled amplitude modulator for the incident signal $E_{d}$. According to the above analysis, we point out that the simultaneous amplification or attenuation can not occur for the two signals $E_{s}$ and $E_{d}$.

In conclusion, we have presented an accessible scheme for selectively implementing controllable microwave amplification and attenuation in a cyclic three-level $\triangle$-type superconducting quantum circuit. Such a project has been implemented by the following two steps: (1) by driving the $\triangle$-type artificial system with three incoming waves, two reverse three-wave mixing processes, sum- and difference-frequency generation with second-order nonlinearity, can coexist; (2) by designing two subtle relations among three incoming waves, interference effects between the incoming signals and the generated signals work and play a crucial role in amplifying or attenuating two output signals.  As the interference terms are sensitive to the relative phase and the driving-field frequency, we can selectively obtain output amplification and attenuation by adjusting these two parameters. Thus the mentioned superconducting system can act as a multifunctional amplitude modulator. This device may have potential applications in solid-state quantum information technology, such as optical switch, quantum feedback, and photon blockade. A current promising candidate for its experimental demonstration is a superconducting fluxonium quantum circuit. Quantum physics underlying the absence of selection rules can be explored further with the aid of superconducting circuits and progresses in this area would open up interesting new avenues for future research and applications. For example, in a recent work quantum routing of single photons with two output channels has been investigated using a cyclic three-level system~\cite{Zhou}.

This work was partially supported by the National Natural Science Foundation of China under the Grant No. 11274132.

\end{document}